\documentclass[12pt,preprint]{aastex}
\usepackage{graphicx}

\title{Misalignment of the Jet and the Normal to the Dusty Torus in the Broad Absorption Line QSO FIRST\,J155633.8+351758}
\author{Cormac Reynolds\altaffilmark{1}, Brian Punsly\altaffilmark{2}, Christopher P. O'Dea\altaffilmark{3}}
\altaffiltext{1}{ICRAR-Curtin University, GPO Box U1987, Perth,
Western Australia, 6102, Australia} \altaffiltext{2}{1415 Granvia
Altamira, Palos Verdes Estates CA, USA 90274 and ICRANet, Piazza
della Repubblica 10 Pescara 65100, Italy, brian.punsly1@verizon.net
or brian.punsly@comdev-usa.com} \altaffiltext{3}{Laboratory for
Multiwavelength Astrophysics, School of Physics and Astronomy,
Rochester Institute of  Technology, 54 Lomb Memorial
Drive,Rochester, NY 14623}
\begin{document}
\begin{abstract}
We performed VLBA observations of the Broad Absorption Line Quasar
FIRST\,J155633.8+351758, ``the first radio loud BALQSO". Our
observations at 15.3~GHz partially resolved a secondary component at
positional angle (PA) $\approx 35^{\circ}$. We combine this
determination of the radio jet projection on the sky plane, with the
constraint that the jet is viewed within $14.3^{\circ}$ of the line
of sight (as implied by the high variability brightness temperature)
and with the position angle (PA) of the optical/UV continuum
polarization in order to study the quasar geometry. Within the
context of the standard model, the data indicates a ``dusty torus"
(scattering surface) with a symmetry axis tilted relative to the
accretion disk normal and a polar broad absorption line outflow
aligned with the accretion disk normal. We compare this geometry to
that indicated by the higher resolution radio data, brightness
temperature and optical/UV continuum polarization PA of a similar
high optical polarization BALQSO, Mrk\,231. A qualitatively similar
geometry is found in these two polar BALQSOs; the continuum
polarization is determined primarily by the tilt of the dusty torus.
\end{abstract}

\section{Introduction}
Approximately 15\% to 20\% of quasars have UV broad absorption line
(BAL) outflows or winds, depending on the sample selection criteria
\citep{hew03,rei03,tru06,gib09}\footnote{We use the original
definition of a BAL as UV absorbing gas that is blue shifted at
least 5,000 km/s relative to the QSO rest frame and displaying a
spread in velocity of at least 2,000 km/s, \citep{wey91}. Note that
this definition specifically excludes the so-called ``mini-BALQSOs,"
with the BALNicity index = 0 \citep{wey97}. This is desirable since
the DR5 statistical analysis of \citet{zha10} indicate that these
types of sources (mini-BALQSOs have a large overlap in definition
with the intermediate width absorption line sources of
\citet{zha10}) tend to resemble non-BALQSOs more than BALQSOs in
many spectral properties. Mini-BALQSOs also tend to have much
smaller X-ray absorbing columns than BALQSOs \citep{pun06}}. The
wind is believed to be a general feature in quasars, but it is seen
only in favorable configurations due to a limited solid angle
\citep{wey97}. The physical origin of the BAL wind is very
uncertain. Knowing the orientation of the BAL wind relative to the
accretion disk normal and dusty torus normal (within the ``standard
quasar model," see Antonucci 1993) would advance our knowledge of
the physics of the BAL wind launching mechanism.

\par Gathering evidence of quasar orientation is difficult.
Most of our knowledge derives from studies of radio loud quasars
\citep{orr82}. Unfortunately, radio jets in BALQSOs are suppressed
and especially so on scales larger than a few kpc, a trend that
strengthens with increasing BALnicity index \citep{bec00,bec01}.
Motivated by the potential orientation information, studies of the
rare subclass of radio loud BALQSOs have been an active area of
research. Statistical studies have been aimed at exploring the
evidence for evolutionary models \citep[e.g.][]{bri84, san02} in which the
BALQSO phenomenon is a transient stage or models based on a preferred line of
sight to an ever-present BAL wind. The orientation models include various wind
geometries: equatorial, \citet{mur95}; mid-latitude, \citet{elv00}, polar
\citet{gho07} or polar winds coexisting with equatorial winds \citep{pun00}.
The radio surveys of \citet{bru12,dip11,mon08} have drawn no clear conclusions.

\par Only recently has direct information on this geometric
configuration become available from radio observations. If a source
is highly variable, one might infer a brightness temperature,
$T_{B}> 10^{12}$ K, that requires a nearly pole-on orientation and
relativistic motion for the jet in order to avoid the ``inverse
Compton catastrophe" \citep{mar79}. Numerous BALQSOs in a pole-on
orientation have now been found by this method
\citep{zho06,gho07,rey09}. The extra constraint provided by the
polar orientation makes these BALQSOs valuable laboratories for
studying quasar geometry. The position angle (PA) on the sky plane
and kinematics of the radio jet provide probes of both the
enveloping BAL wind and the geometry of the wind/accretion disk
system. Another constraint on the quasar geometry is provided by the
optical continuum polarization PA. This relates directly to the
geometry of the UV absorbing wind and, even more so, the scattering
surface that is associated with the elevated polarization seen in
some BALQSOs. Thusly motivated, we have begun a program to study
subparsec and parsec scale radio jets in combination with optical
polarization PAs in BALQSOs.

\par The Doppler enhancement associated with the polar
orientation tends to make the jets in polar BALQSOs amongst the most
luminous in a class of objects that have a strong propensity for
weak jet emission\footnote{Typically, VLBI ``BALQSO" targets are
actually mini-BALQSOs since they have larger radio fluxes than
bona-fide BALQSOs, e.g. \citet{bru13, hay13}}. In order to pursue a
study of polar BALQSO geometry, high resolution (high frequency),
VLBI (Very Long Baseline Interferometry) observations with high
sensitivity (because of the weak secondary components and jets) are
required in order to see the jet direction near the accretion disk.
Mrk\,231 is the best choice. The proximity, $z=0.042$, and the large
flux density of $\sim 100$ mJy at 22~GHz are two factors that
greatly enhance the spatial resolution achievable in observations.
We have examined high frequency VLBA (Very Long Baseline Array)
radio observations of Mrk\,231 and their geometric implications
\citep{rey09}. It is essential to do the same for more BALQSOs in
order to understand which geometric properties are endemic to the
broad absorption phenomenon and which properties are unrelated
idiosyncrasies of particular sources. This critical aspect
challenges the limits of our observational tools since all other
bona-fide BALQSOs are much fainter radio sources than Mrk\,231. The
next best choice after Mrk\,231 is FIRST\,J155633.8+351758
(FIRST\,J1556+3517, hereafter), the high optical polarization quasar
at z=1.50, dubbed the first discovered radio loud BALQSO
\citep{bec97}. Strangely, like Mrk\,231, this is a member of the
rare class of BALQSOs known as FeLoBALQSOs due to absorption in low
ionization species and strong Fe II emission. Being the ``first"
radio loud BALQSO discovered led to a flurry of deep optical
observations \citep{bro97, naj00}. The polarization PA $\approx
153^{\circ}$ from deep Keck spectroscopy and it has a flat 25 to 30
mJy spectrum from 1.4 to 5 GHz \citep{bro97,bec97}.
FIRST\,J1556+3517 is viewed close to the polar axis ($<
14.3^{\circ}$) based on high variability brightness temperatures
\citep{gho07}.

\par The paper is organized as follows. In Section~2, we discuss our VLBA
observations ranging from 1.6 GHz to 15.3 GHz and the implications
for the quasar geometry. Section~3 is a consideration of the
geometrical implications of the observations. The next section
discusses the implicit geometry in a broader context that includes
Mrk\,231. Throughout this paper, we adopt the following cosmological
parameters: $H_{0}$=71 km/s/Mpc, $\Omega_{\Lambda}=0.73$ and
$\Omega_{m}=0.27$.

\section{The VLBA Observations} A previous VLBI observation of
FIRST\,J1556+3517 was unresolved with a flux density of 26~mJy with
the EVN at 1.4~GHz \citep{jia03}. Better resolution was needed to
detect a jet, so we carried out observations at frequencies of 1.6,
5.0, 8.4 and 15.3~GHz with the VLBA at epoch 2011.07 (project code
BR156) -- see Table~\ref{table:observations}. A recording rate of
512~Mbps (the best available at the time) was used in order to
achieve the high sensitivity required to detect a weak jet. This
provided 64 MHz of bandwidth in 2 polarizations with 2-bit Nyquist
rate sampling at each frequency. The resultant image sensitivities
were 0.11, 0.08, 0.08 and 0.15 mJy/beam at 1.6, 5.0, 8.4 and
15.3~GHz respectively. The data were correlated on the VLBA
correlator in Socorro, NM and calibration was carried out in the
standard way using NRAO's Astronomical Image Processing System
(AIPS) package via the ParselTongue interface \citep{ket06}.
FIRST\,J1556+3517 was phase referenced to the nearby calibrator
J1602+3326 to provide phase calibration and some astrometric
information. Subsequent phase self-calibration of the target source
was also possible at all frequencies.

\begin{figure}
\includegraphics[width= 80 mm]{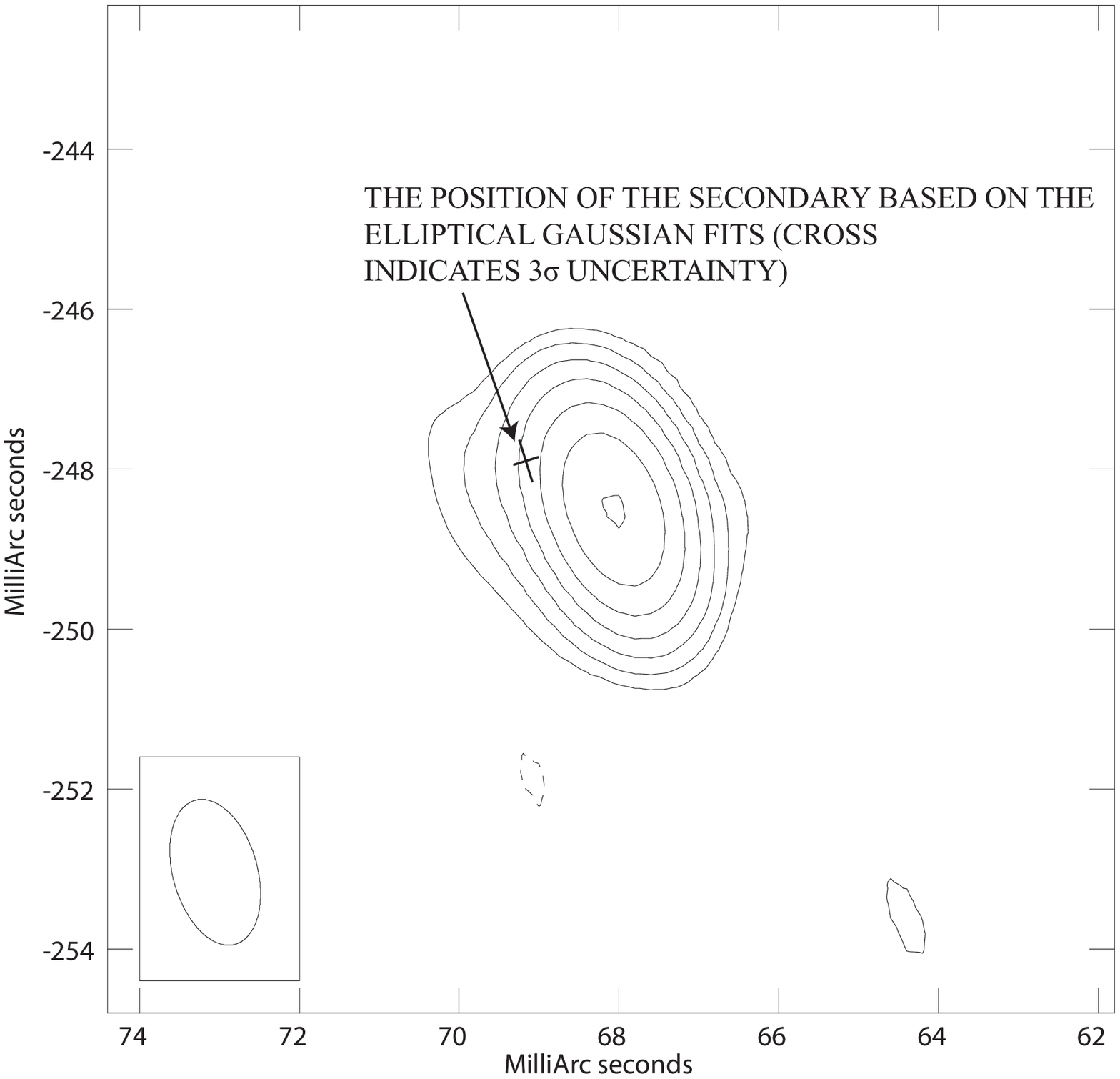}
\includegraphics[width= 80 mm]{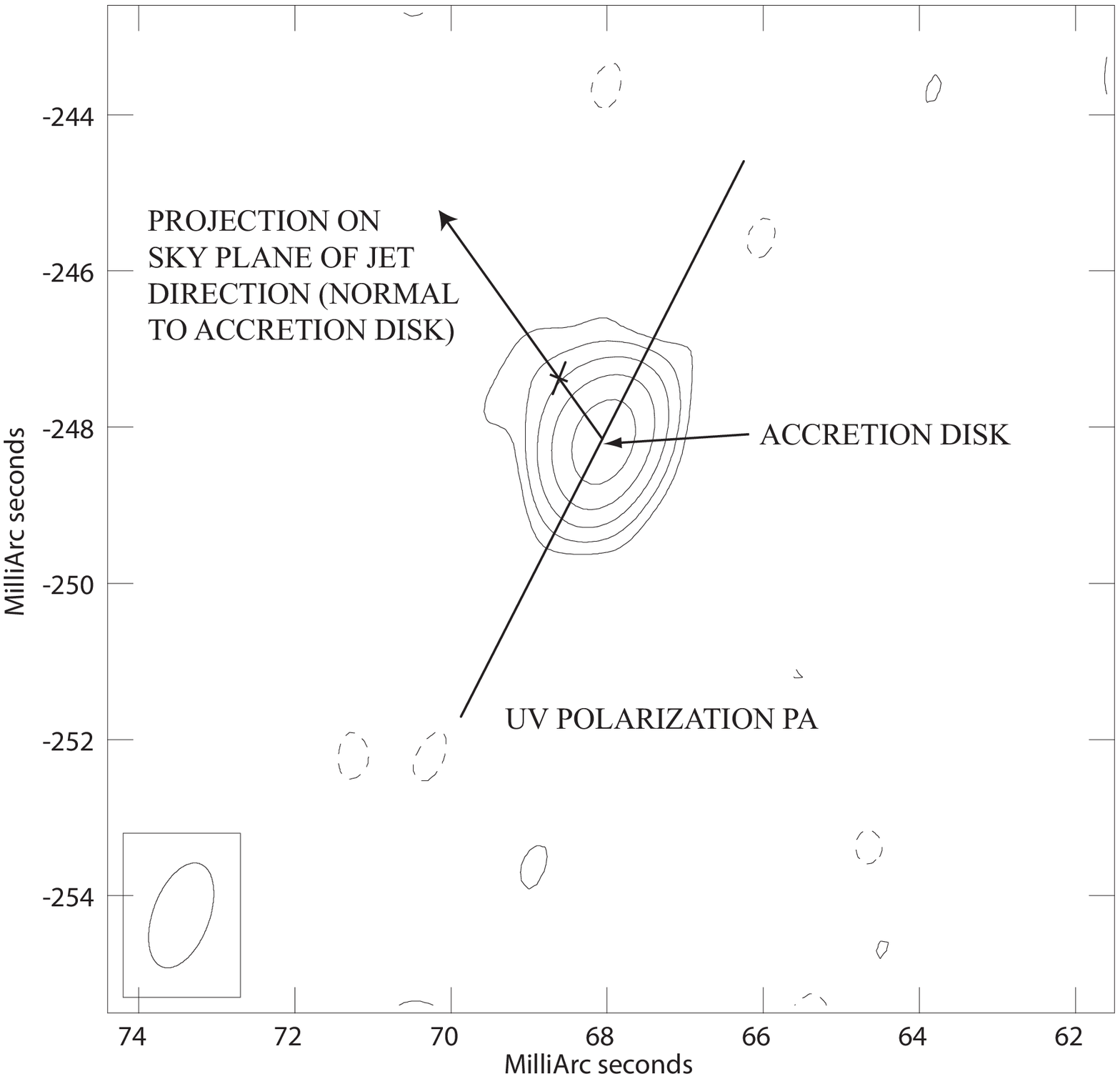}
\caption{On the left (right) is an 8.4 GHz (15.3 GHz) image of
FIRST\,J1556+3517. Contours start at $3\sigma$ above the rms image
noise. The 8.4 GHz contours start at $\pm 0.23$ mJy/beam and
increase by factors of two to a peak of 14.9 mJy/beam. The 15.3 GHz
contours start at $\pm 0.45$~mJy/beam and increase by factors of two
to a peak of 11.2 mJy/beam. The position of the secondary jet
component is indicated by the small crosses (the sizes of which
indicate the $3\sigma$ position uncertainty along the major and
minor axes of the restoring beam). The restoring beams are indicated
in the bottom left insets. The right hand frame shows the relative
orientation of the projection of the jet in the sky plane inferred
from the secondary position and the optical/UV polarization PA.}

\end{figure}

\section{Results and Geometric Implications}
At 1.4 and 5 GHz, the source was unresolved with synthesized beam
sizes of approximately $7 \times 10$~mas and $2 \times 3$~mas
respectively. As with previous VLA and EVN observations, there is no
evidence of jet re-orientation in these images in the form of
resolved emission at a differing position angle from the VLBI
secondary component. However, at 8.4 and 15.3~GHz, with synthesized
beams of $1.0 \times 1.8$ and $0.7 \times 1.1$~mas respectively, we
partially resolve a faint secondary component (see Figure~1)
allowing us to determine the direction of the parsec-scale jet. We
fitted a two Gaussian component model to the 8.4 and 15.3~GHz data
using the DIFMAP software package \citep{shepherd95}, the results of
which are given in Table~\ref{table:observations}. The positions of
the 8.4 GHz and 15.3 GHz secondary components do not agree, likely a
combination of inhomogeneous optical depth effects and a curving
jet. The 15.3 GHz observation, sensitive to emission originating
much deeper within the optically thick plasma, is more relevant for
exploring nuclear geometry.
\begin{table}
\caption{Parameters of Core and Jet Components in FIRST\,J1556+357 at Epoch 2011.07
\label{table:observations}
}
\begin{tabular}{ccccc} \tableline \rule{0mm}{3mm}
Frequency/(GHz)       & Core Flux      & Secondary Flux & Separation (mas) & Secondary (PA) \\
                      & Density/(mJy)  & Density/(mJy)  &                  & Degrees        \\
\hline

1.6 \tablenotemark{a} & 16.5 $\pm 0.5$ & --             & --               & --             \\
5.0 \tablenotemark{a} & 18.8 $\pm 0.5$ & --             & --               & --             \\
8.4                   & 17.2 $\pm 0.5$ & 0.7 $\pm 0.1$  & $1.57 \pm 0.09$  & $61 \pm 3$     \\
15.3                  & 14.1 $\pm 0.5$ & 0.9 $\pm 0.1$  & $0.98 \pm 0.08$  & $35 \pm 3$     \\
\tableline \rule{0mm}{3mm}

\end{tabular}
\tablenotetext{a}{core and secondary not resolved}
\end{table}

The VLBA observations, in conjunction with the UV polarization PA
(position angle), provide a set of tight constraints on the quasar
geometry (the right hand frame of Figure~1). These can be summarized
as
\begin{enumerate}
\item
The accretion disk position is associated with the high frequency
radio core (right hand frame Figure~1).
\item In \citet{gho07}, the variability brightness temperature was calculated as $T_{B}= 6.7
\times 10^{13}\,\mathrm{K}$, thereby restricting the jet axis to $<
14.3^{\circ}$ to the line of sight.
\item
The projection of the radio jet axis on the sky plane, computed from
the coordinates of the peaks of Gaussian components, is at PA
$\approx 35^{\circ}$ (right hand frame Figure~1).
\item
From 2 and 3 above, the accretion disk normal (parallel to the pc
radio jet) is rotated relative to the z axis (the line of sight
which comes out of the page) by $< 8.2^{\circ}$ about the vertical,
y-axis (all rotations are in a left handed sense) and $<
11.7^{\circ}$ about the horizontal, x-axis (which points to the
right), see Figure~2. The direction of the jet in Figure~2
represents the maximum angle from the line of sight.
\item
The UV polarization direction is $\textrm{PA} \approx 153^{\circ}
\equiv -27^{\circ}$ redward of $3000$~\AA\@ (see Figure~2)
\citep{bro97}.
\item
The scattered photons are along the line of sight by definition
(the z-axis).
\item
Since the line of sight is through the BAL wind and the jet is aligned close to the line of
sight, the BAL wind is a polar outflow aligned close to the line of sight.
\end{enumerate}

\par Combining these constraints, we deduce the relative orientation of the
scattering surface that produces the high polarization and the BAL
wind. Since the emission lines are polarized identically to the
continuum, \cite{bro97} pointed out that the scattering surface lies
well outside the broad emission line region, $\sim 1$ pc from
the source. This is a typical distance to the inner edge of the
dusty torus \citep{bar87}. A scatterer containing some dust seems
reasonable based on the reddened spectrum of the polarized emission
\citep{naj00}. An asymmetric distribution of dusty scatterers has
been routinely invoked to explain the similar polarization
properties of another reddened BALQSO, Mrk\,231, which also has a
polarization level that rises rapidly into the near UV
\citep{smi95,goo94,sch85,tho80}. For example, within the context of
the ``standard model" we may have a torus with a symmetry axis that
is tilted relative to the jet axis by $30^{\circ}$ to $40^{\circ}$
by means of a rotation about a line that is parallel to the PA of
the projection of the polarization on the sky plane, i.e., a line at
$PA = 153^{\circ}$ in the x-y plane (see Figure~2). Misaligned tori
are not an expected feature of the standard model. However, direct
evidence of misalignment of molecular gas on subparsec and parsec
scales exists. There are 10 AGN with water maser disks (viewed edge
on) that have been observed which also have high resolution images
of the radio jets \citep{gre13,kon05}. One of these AGN (NGC 1068)
also has Mid-IR interferometry on parsec scales that are typical for
the dusty torus \citep{rab09}. Even though the statistics are small,
the data indicates a significant misalignment of the normal to the
plane of molecular gas rotation and the subparsec or parsec scale
jet of $\geq 30^{\circ}$ in 20\% to 30\% of AGN. So the occurrence
of a misaligned torus in some BALQSOs should not be unexpected. The
angle of misalignment, although speculative, is motivated in the
next section.

\begin{figure}
\begin{center}
\includegraphics[width= 0.8\textwidth ]{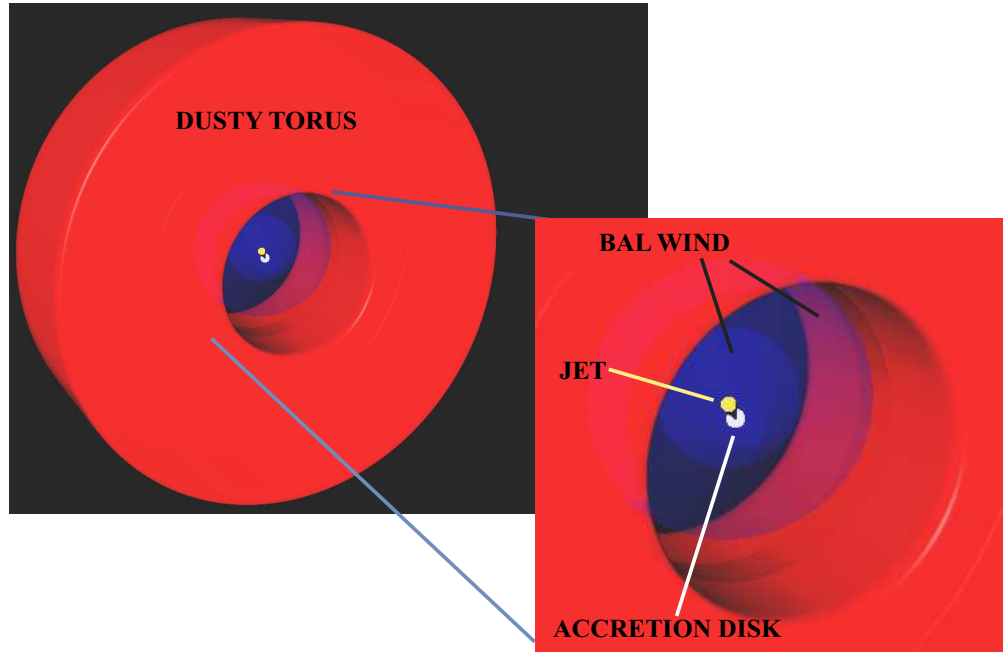}
\caption{A simple geometry for FIRST\,J1556+3517 that is consistent
with the VLBA observations and the Keck polarization data. A dusty
torus in red harbors a small white (almost face-on) accretion disk
at its centre. Emanating from the accretion disk is a narrow radio
jet that is depicted by a cone with black sides and a yellow cap. It
is beamed almost directly towards Earth, but slightly to the North
and East. It is nested within a coaxial diffuse light blue BAL
conical wind. The density gradient in the conical wind is indicated
schematically by a high opacity blue cone nested within a low
opacity light blue cone. The tilt of the torus in the sky plane
determines the polarization PA. On the right is a zoom in of the
central region.}
\end{center}
\end{figure}
\begin{figure}
\includegraphics[width= 73 mm]{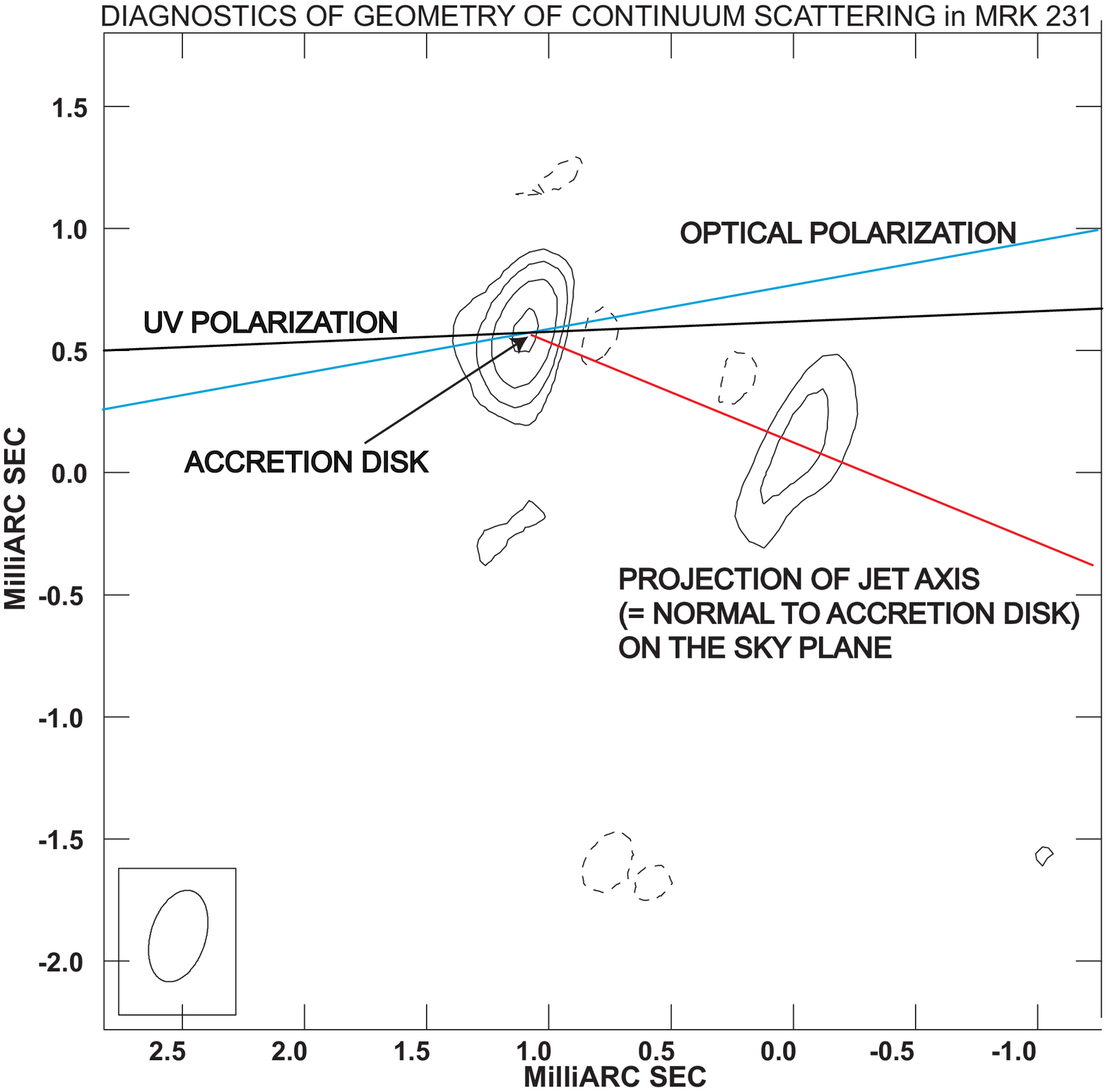}
\hspace{0.8cm}
\includegraphics[width= 79 mm]{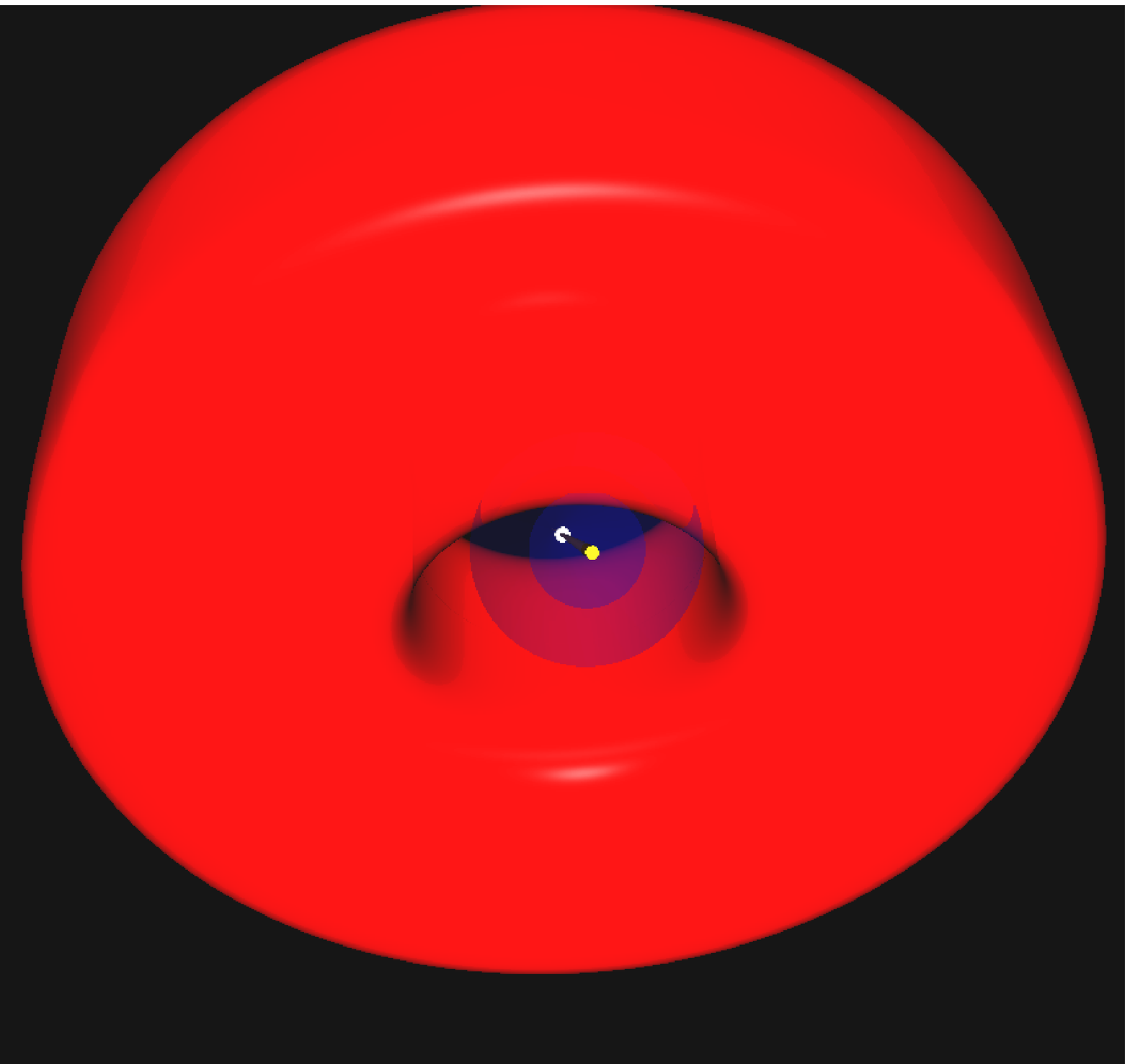}
\caption{The left frame shows a secondary 0.8 pc from radio core
in this 43~GHz VLBA image of Mrk\,231 from \citet{rey09}. The optical
and UV continuum polarization directions from \citet{smi95} are
superimposed on the radio image. The left hand frame shows the
relative orientation of the projection of the jet in the sky plane
inferred from the secondary position and the optical/UV polarization
PA. The right hand frame depicts a 3-D physical representation of
the 2-D data projected on the sky plane that is presented in the
left frame.}
\end{figure}

FIRST\,J1556+3517 is a very high polarization BALQSO with 12\%
continuum polarization in the near UV \citep{bro97}. A misalignment
of the dusty torus and the radio jet is expected in high
polarization ($>3\%$) BALQSOs based on theoretical modeling of
polarization in polar BAL wind models \citep{pun00}. Theoretically,
large attenuation of the continuum is achieved in LoBALQSOs by the
base of the BAL wind (electron scattering) and dusty gas that is
entrained in the wind on parsec scales. Thus, there is a large
contribution of scattered light to the net observed flux. Since the
BAL wind is parallel to the jet and the line of sight, in order for
high polarization to occur in polar BALQSOs, the symmetry of the
system must be broken. This is either accomplished by anisotropic
scatterers (see Kim and Martin 1995) or, as considered more likely
here, an asymmetric distribution of isotropic scatters
\citep{bro77}. If the large polarization is from anisotropic
scatterers then the polarization direction might be determined by
dust grain alignment in a large scale wind or a large scale ordered
magnetic field. Although this alternative scattering scenario is not
ruled out by observation, it is not pursued here.

\section{Analogies to Mrk\,231}
In order to see if this scenario of an asymmetric distribution of isotropic
scatters is reasonable, it would be instructive to compare our findings to
another similar object. Mrk\,231 is also a FeLoBALQSO with high near UV
continuum polarization, $\sim 13\%$ \citep{smi95}. The data for Mrk\,231 is
superior due to its proximity to Earth. In \citet{rey09}, we considered two
epochs of 43, 22 and 15 GHz VLBA observations of the resolved core of Mrk\,231
and also the time variability of the historic light curve at 22 GHz. A resolved
secondary exists just 0.8 pc from the core (see Figure~3). Repeating the
evidence and logic that we used to determine a nuclear geometry for
FIRST\,J1556+3517 in Section~3 with the data gathered from observations of
Mrk\,231 we find:

\begin{enumerate}
\item The accretion disk position is associated with the radio core at 43 GHz (see Figure~3, left frame).
\item The variability brightness temperature, $T_{B}= 1.25 \times 10^{13}\,\mathrm{K}$,
restricts the jet axis to $< 25.6^{\circ}$ to the line of sight \citep{rey09}.
\item The projection of the radio jet axis on the sky plane is at PA
$\approx -112^{\circ}$ (see Figure~3, left frame).
\item From items 2 and 3 above, the accretion disk normal (parallel to the
parsec radio jet) is rotated relative to the z axis (the line of
sight which is out of the page) by $< 23.5^{\circ}$ about the
vertical, y-axis (all rotations are in a right handed sense) and $<
10.1^{\circ}$ about the horizontal, x-axis (which points to the
right). The angle of the jet in Figure~3 represents the maximum
angle from the line of sight.
\item The polarization direction from HST and ground based observations in \citet{smi95} is $90^{\circ} < \textrm{PA} < 100^{\circ}$
redward of $3000 \AA$ (as depicted in the left hand frame of Figure~3).
\item The scattered photons are along the line of sight by definition
(the z-axis).
\item Since the line of sight is through the BAL wind and the jet is aligned close to the line of
sight, the BAL wind is a polar outflow aligned close to the line of
sight.
\end{enumerate}

\par The right hand frame of Figure~3 is a realization of the polarization and
radio data in terms of the components of the standard model. A
tilted dusty torus exists as for FIRST\,J1556+3517. The polar BAL
wind directed close to the line of sight simplifies the scattering
geometry. The scattering mirror is essentially the small unshadowed
region of the surface that bounds the hole of the torus almost due
South. The projection of the tangent plane at the ``centroid" of
this effective mirror, as seen in the sky plane, is almost parallel
to the PA of the continuum polarization (reflected light is
polarized perpendicular to the scattering plane). It would be
exactly parallel to the PA of the continuum polarization if the BAL
wind were viewed exactly pole-on. The angle of the tilt was found by
rotating the angle of the torus until the desired level of
polarization was obtained (Punsly et al. in preparation). This result
is assumption dependent and to match the polarization and
attenuation in the UV requires the entrainment of exotic dust in the
BAL wind, the type of non-Galactic dust that is seen in some distant
galaxies in the study of gravitational lenses and supernovae
\citep{eli06,fol10}. The large attenuation in the UV requires small
grain sizes \citep{kim95}, and we assume them to be isotropic
scatterers. The dust entrained in the BAL wind attenuates both the
continuum and the light reflected from the torus, but does it not
polarize the emission that it scatters similar to what is seen in
the centers of cold dark clouds in the Galaxy \citep{goo95}. The
large number of free parameters in the models detracts greatly from
their predictive power.

\section{Conclusion}
In this letter, we demonstrated how $T_{B}$, the optical continuum
PA and the radio jet direction can be used to determine a compatible
geometry for individual BALQSOs within the context of the
fundamental elements of the standard model. This was illustrated
with two similar high polarization sources -- FIRST\,J1556+3517 and
Mrk\,231, indicating a misaligned accretion disk and dusty torus.
Higher frequency and higher sensitivity VLBA observations of
FIRST\,J1556+3517 (using the new wideband digital backend system)
will allow us to better determine the direction of the jet on even
smaller scales and better constrain the geometry (recall the
evidence for a curving jet discussed in Section~3). If higher
resolution observations indicate a change to the estimate of the jet
PA then the methods presented here are robust and the geometric
picture in Figure~2 (the jet direction) can be modified in a
straightforward manner. Furthermore, multi-epoch monitoring of
superluminal motion of secondaries in FIRST\,J1556+3517 and Mrk\,231
can supplement the $T_{B}$ estimates in our determination of the
angle between the line of sight and the jet direction.

\section{Acknowledgments}
The National Radio Astronomy Observatory is a facility of the National Science Foundation operated under cooperative agreement by Associated Universities, Inc. This work made use of the Swinburne University of Technology software correlator, developed as part of the Australian Major National Research Facilities Programme and operated under licence.

\end{document}